\documentclass[aps,pra,twocolumn,showpacs, superscriptaddress]{revtex4-1}

\usepackage{graphicx}
\usepackage{color}
\usepackage[usenames,dvipsnames]{xcolor}
\usepackage{amsmath,amsthm,amsfonts,amssymb,bm}

\newcommand{\nc}{\newcommand}
\nc{\eq}{\begin{equation}}
\nc{\eeq}{\end{equation}}
\nc{\eqa}{\begin{eqnarray}}
\nc{\eeqa}{\end{eqnarray}}
\nc{\ar}{\begin{array}}
\nc{\ear}{\end{array}}
\nc{\bfig}{\begin{figure}}
\nc{\efig}{\end{figure}}
\nc{\dg}{\dagger}
\nc{\eps}{\frac{\epsilon}{2}}
\nc{\juuri}{\sqrt{\Omega^2+(\eps)^2}}
\nc{\sx}{\sigma_x}
\nc{\sy}{\sigma_y}
\nc{\sz}{\sigma_z}
\nc{\spl}{\sigma_+}
\nc{\sm}{\sigma_-}
\nc{\Sx}{\bar{\sigma}_x}
\nc{\Sy}{\bar{\sigma}_y}
\nc{\Sz}{\bar{\sigma}_z}
\nc{\Spl}{\bar{\sigma}_+}
\nc{\Sm}{\bar{\sigma}_-}
\nc{\nn}{\nonumber}
\nc{\noi}{\noindent}
\nc{\omt}{\tilde{\omega}}
\nc{\Somt}{S(\omt)}
\nc{\Somtd}{S^{\dg}(\omt)}
\nc{\got}{\gamma_{\omega}(t)}
\nc{\gmot}{\gamma_{-\omega}(t)}
\nc{\po}{\mathcal{P}}
\nc{\qo}{\mathcal{Q}}
\nc{\adg}{a^{\dg}}
\nc{\gammat}{\tilde{\gamma}}
\nc{\Q}{$\mathcal{Q }$}
\nc{\C}{$\mathcal{C }$}
\nc{\kvec}{\mathbf{k}}

\def\bra#1{\mathinner{\langle{#1}|}}
\def\ket#1{\mathinner{|{#1}\rangle}}

\begin{document}

\title{Time-Invariant Discord in Dynamically Decoupled Systems}

\author{Carole Addis}
\email{ca99@hw.ac.uk}
\affiliation{SUPA, EPS/Physics, Heriot-Watt University, Edinburgh, EH14 4AS, UK}
\author{G\"{o}ktu\u{g} Karpat}
\affiliation{Turku Center for Quantum Physics, Department of Physics and Astronomy, University of Turku, FIN-20014, Turun yliopisto, Finland}
\affiliation{Faculdade de Ci\^encias, UNESP - Universidade Estadual Paulista, Bauru, SP, 17033-360, Brazil}
\author{Sabrina Maniscalco}
\email{smanis@utu.fi}
\affiliation{Turku Center for Quantum Physics, Department of Physics and Astronomy, University of Turku, FIN-20014, Turun yliopisto, Finland}

\date{\today}

\begin{abstract}
Dynamical decoupling protocols are one of the most used tools for efficient quantum error corrections and for reservoir engineering.
In this paper we study the effect of dynamical decoupling pulses on the preservation of both quantum and classical correlations, and their influence on the intriguing phenomenon of time-invariant discord. We study two qubits experiencing local dephasing with an Ohmic class spectrum and subject to dynamical decoupling protocols. We investigate the connection between the dynamics of both classical and quantum correlations and the behaviour of information flow between system and environment. Finally, we establish a set of necessary conditions for which time-invariant discord can be created using dynamically decoupling techniques.
\end{abstract}


\maketitle

\section{Introduction}

Genuine quantum traits possessed by quantum mechanical systems, such as coherence and quantum correlations, are fundamental for all possible applications of quantum information science. Realistic quantum systems are however subjected to the interaction with their surroundings, which consistently results in the loss of such critical quantum features \cite{book}. Since quantum devices need to be robust against the destructive effects of their environment to be used in everyday tasks, protecting coherence and quantum correlations as long as possible is both of great importance and a major challenge.

Several methods have been introduced to suppress the noise induced on the system by the environment in order to preserve quantum correlations. One such strategy, known in the literature as dynamical decoupling (DD), relies on the application of a sequence of external pulses to the principle system so that the harmful effects of the environment are countered \cite{dd1,dd2,dd3,dd4,dd5}. DD techniques for open quantum systems are among the most successful methods to subdue decoherence. Specifically, ``bang bang" periodic dynamical decoupling (PDD) schemes can be exploited to prolong coherence times and restore monotonically decaying correlations in quantum systems which are undergoing decoherence \cite{dd6,addis}.

Another approach to counteract the environmental noise is based on the exploitation of memory effects characterising non-Markovian dynamics \cite{nmreview1,nmreview2}. While in Markovian open  systems information is monotonically lost to the environment, non-Markovian dynamics exhibits backflow of information from the environment to the system. This in turn results in the emergence of memory effects since, as a consequence of the backflow, future states of the system might depend on its past states. Thus, similarly to DD techniques, non-Markovianity can also be used to restore monotonically decaying correlations \cite{bellomo}. An ongoing debate concerns itself with characterizing and quantifying non-Markovianity \cite{rivas10,hou11,breuer09,lu10,luo12,fanchini14,bylicka14,haseli14,chr14}. At present, no universal measure exists and indeed in some cases, connections between measures can not be established  \cite{notagree1,notagree2}. However, our treatment here is not quantitatively dependent on a specific measure, allowing our results to be presented independently of a specific quantifier of non-Markovianity.

Entanglement has been considered as the main resource of quantum protocols for a long time \cite{entreview}. However, it has been later on demonstrated via numerous examples that quantum discord \cite{discordreview} is also essential for, e.g., distribution of entanglement \cite{discorduse1}, quantum locking \cite{discorduse2}, entanglement irreversibility \cite{discorduse3} and many other tasks \cite{discorduse4}. Quantum discord has been shown to be frozen at a finite value for a certain time interval \cite{discfro} or even forever \cite{discinv}, depending on the properties of the initial state and the specifics of the interaction with the environment. In particular, for two qubits in Bell-diagonal states and independently interacting with dephasing environments, while frozen discord manifests itself for both Markovian \cite{discfro} and non-Markovian \cite{discfronm} environments, the emergence of time-invariant discord is more closely related to the presence of memory \cite{discinv}.

The use of either DD techniques or non-Markovian effects prolongs the preservation of both entanglement and discord in presence of environmental noise  \cite{1,2,3,4,5}. However, studies of single qubit dynamics have recently shown that the simultaneous use of non-Markovian reservoir engineering and DD protocols is counterproductive for coherence preservation \cite{addis}. Motivated by this rather unexpected result, here we study the possibility of creating time-invariant discord through PDD techniques in relation to the non-Markovian behavior in the dynamics. A clear merit in creating time-invariant discord is in the execution of quantum protocols based on discord, which become unaffected by noise under the specific conditions. In this work, we consider two qubits interacting with a dephasing environment having an ohmic type spectral density, focussing on the cases of both one-sided and two-sided local noise. Our investigation allows us to establish a set of necessary conditions for which time-invariant discord can be observed.

The paper is structured as follows. Section II introduces the system, namely a pure dephasing model, along with its exact solution in presence of DD. In Section III, we state the main results of our study, i.e., we present a detailed investigation of the phenomenon of time-invariant discord and determine a set of necessary conditions for which time-invariant discord manifests when the system is periodically pulsed. In Section IV we present conclusions.

\section{The Model}

We consider a microscopic model describing the dephasing interaction of two identical qubits with their respective local independent environments represented by a bath of bosonic oscillators. The total Hamiltonian of each of the two independent qubit-environment parts can be written as follows
\eq
H=\omega_0\sigma_z+\sum_k\omega_k a^\dagger_k a_k+\sum_k\sigma_z(g_ka_k+g_k^*a_k^{\dagger}),
\eeq
in units of $\hbar$. Here, $\sigma_z$ is the $z$-component of the spin operator for the two-level systems, with $\omega_0$ the qubit transition frequency and $\omega_k$ the frequency of the $k$th reservoir mode. The corresponding annihilation (creation) operators are given by $a_k (a_k^\dagger)$, and $g_k$ is the coupling constant associated with the qubit-$k$th mode interaction. This model is exact and admits an analytical solution \cite{puredephasing1,puredephasing2,puredephasing3}. In the interaction picture, the master equation for the density matrix of the qubit is given by
\eq
\dot\rho=\gamma_0(t)[\sigma_z\rho\sigma_z-\rho]/2,
\eeq
The resulting decay of the off-diagonal density matrix elements are as follows: $\rho_{01}(t)=\rho_{10}^*(t)=\rho_{01}(0)e^{-\Gamma_0(t)}$ where $\gamma_0(t)=d\Gamma_0(t)/dt$ and
\eq
\Gamma_0(t)=\int_0^{\infty}\frac{I(\omega)}{\omega^2}[1-\cos(\omega t)]d\omega,
\label{eq:Gamma0}
\eeq
where $I(\omega)=\sum_j\delta(\omega-\omega_j)|g_j|^2$ is the spectral density function characterising the interaction of the qubit with the oscillator bath, which is assumed to be initially at zero temperature. In our work, we consider the widely studied class of spectral densities of the form
\eq
I(\omega)=\frac{\omega^s}{\omega_c^{s-1}}e^{-\omega/\omega_c}\label{SDohm}
\eeq
with $s$  being the Ohmicity parameter and $\omega_c$ a cutoff frequency. Ohmic spectra correspond to the case $s=1$, sub-Ohmic to $s<1$ and super-Ohmic to $s>1$.

Next, we introduce the analytical solution of the above dephasing model in the presence of an arbitrary sequence of bang-bang pulses modelled as instantaneous $\pi$-rotations about the $x$-axis. In this situation, the reduced dynamics of the qubit can still be exactly described by exchanging $\Gamma_0(t)$ with a modified decoherence function $\Gamma(t)$. Suppose that we consider an arbitrary storage time, $t$, during which a total number of $N$ pulses are applied at instants $\{t_1,...t_n,...t_f\}$, with $0<t_1<t_2<...<t_f<t$. Then, the decoherence function $\Gamma(t)$ can be expressed as \cite{uhrig}
\eq
\Gamma(t)= \left\{
  \begin{array}{l l}
    \Gamma_0(t) & t\leq t_1 \\
    \Gamma_n(t) & t_n<t\leq t_{n+1},0<n<N \\
    \Gamma_N(t) & t_f<t \label{eq:Gamma}
  \end{array} \right.,
\eeq
where the exact representation of the controlled decoherence function $\Gamma_n(t)$ for $1\leq n \leq N$, can be written in the following form:
\eq
\Gamma_n(t)=\int_0^\infty \frac{I(\omega)}{2\omega^2}|y_n(\omega t)|^2d\omega, \:\:\:\:\:n\geq0,
\eeq
where $|y_0(\omega t)|^2=|1-e^{i\omega t}|^2$, and 
\eq
y_n(z)=1+(-1)^{n+1}e^{iz}+2\sum_{m=1}^n(-1)^me^{iz\delta_m}, \:\:\:\:\:z>0.
\eeq
Here, it is understood that the $n$th pulse occurs at time $t_n=\delta_nt$ and $0<\delta_1<...<\delta_n<...<\delta_s<1$. In order to express the controlled decoherence function $\Gamma_n(t)$ in terms of its uncontrolled counterpart $\Gamma_0(t)$, we simply relate $|y_n(\omega t)|^2$ to $|y_0(\omega t)|^2$ using iteration to find the following exact expression \cite{dd6}:
\eqa
\Gamma_n(t)&=&2\sum_{m=1}^n(-1)^{m+1}\Gamma_0(t_m)\nn\\&+&4\sum_{m=2}^n\sum_{j<m}\Gamma_0(t_m-t_j)(-1)^{m-1+j}\nn\\&+&2\sum_{m=1}^n(-1)^{m+n}\Gamma_0(t-t_m)+(-1)^n\Gamma_0(t).
\label{Gamma}
\eeqa
In the case of one-sided noise, DD pulses are applied only to one qubit through the modified decay rate $\Gamma(t)$. On the other hand, for two-sided noise, DD pulses are applied to both qubits.

\section{Time-invariant Discord}

Before starting to present our findings, let us first introduce some preliminary concepts and recall several known results form the literature, which have motivated us to investigate the time-invariant discord for dynamically decoupled systems. The total amount of classical and quantum correlations in a bipartite quantum states can be quantified by the quantum mutual information,
\eqa
\mathcal{I}(\rho_{AB})&=&S(\rho_A)+S(\rho_B)-S(\rho_{AB})
\eeqa
where $\rho_{AB}$ and $\rho_{A(B)}$ are the the density matrix of the total system and reduced density matrix of subsystem A(B) respectively and $S(\rho)=-\text{Tr}\{\rho\log_2\rho\}$ is the von Neumann entropy. On the other hand, genuine quantum correlations of a more general type than entanglement are defined as quantum discord \cite{discord1}
\eqa
\mathcal{D}(\rho_{AB})=\mathcal{I}(\rho_{AB})-\mathcal{C}(\rho_{AB}),
\eeqa
where the classical correlations is given by \cite{discord2}
\begin{equation}
\mathcal{C}(\rho_{AB})=\max_{\{\Pi_{k}^{B}\}}\left\{ S\left(\rho_{A}\right)-\sum_{k}p_{k}S\left(\rho_{A|k}, \right)\right\}\label{cc}
\end{equation}
where $\{\Pi_{k}^{B}\}$ is a positive operator valued measure (POVM) acting only on the subsystem $B$, and $\rho_{A|k}=Tr_B(\Pi_{k}^{B}\rho_{AB}\Pi_{k}^{B})/p_k$ is the remaining state of the subsystem $A$ after obtaining the outcome $k$ with probability $p_k=Tr_{AB}(\Pi_{k}^{B}\rho_{AB}\Pi_{k}^{B})$ in the subsystem $B$.

The curious phenomenon of frozen discord has been first realised in Ref. \cite{discfro} for a class of Bell-diagonal states which are locally interacting with Markovian dephasing environments. It was shown that quantum discord might remain constant at a non-zero value despite the effect of the noise for an initial period of time $0<t<\bar t$, during which only classical correlations decay. After the critical time $\bar t$ is reached, as discord is subject to noise-induced decay, classical correlations freeze. Consequently, this unanticipated phenomenon was named as sudden transition between classical and quantum decoherence. On the other hand, for non-Markovian environments, the existence of a transition time $\bar t$ crucially depends not only on the initial state of the bipartite system but also on the specific properties of the spectral density of the environment \cite{discinv}. Indeed, under certain conditions, it was demonstrated that a transition time $\bar t$ does not exist at all and therefore discord remains time-invariant throughout the whole time evolution \cite{discinv}.

As in other works studying time-invariant discord, we focus only on initial Bell-diagonal states of the form
\eq
\rho_{AB}=\frac{(1+c)}{2}\ket{\Phi^{\pm}}\bra{\Phi^{\pm}}+\frac{(1-c)}{2}\ket{\Psi^{\pm}}\bra{\Psi^{\pm}}
\label{state}
\eeq
where $\ket{\Phi}=(\ket{00}\pm\ket{11})/\sqrt{2}$ and $\ket{\Psi}=(\ket{01}\pm\ket{10})/\sqrt{2}$ are the four Bell states and $|c|<1$. We consider two different cases of local dephasing noise in presence of PDD in our work. Firstly, in case where both qubits $A$ and $B$ interact locally with identical dephasing environments having an ohmic type spectral density, the expressions for the mutual information $\mathcal{I}(\rho_{AB})$ and classical correlations $\mathcal{C}(\rho_{AB})$ are given by \cite{luo}
\begin{align}
\mathcal{C}[\rho_{AB}]&=\sum_{j=1}^2\frac{1+(-1)^j\chi}{2}\text{log}_2[1+(-1)^j\chi(t)], \\
\mathcal{I}[\rho_{AB}(t)]&=\sum_{j=1}^2\frac{1+(-1)^jc}{2}\text{log}_2[1+(-1)^jc]\nn\\ &+\sum_{j=1}^2\frac{1+(-1)^je^{-2\Gamma(t)}}{2}\text{log}_2[1+(-1)^je^{-2\Gamma(t)}],
\label{I}
\end{align}
where $\chi(t)=\text{max}\{e^{-2\Gamma(t)},c\}$. Secondly, we let only one of the qubits interact with the dephasing environment and assume that the other qubit is protected from decoherence, in which case the expressions for the mutual information $\mathcal{I}(\rho_{AB})$ and classical correlations $\mathcal{C}(\rho_{AB})$ can be simply obtained by replacing $e^{-2\Gamma(t)}$ with $e^{-\Gamma(t)}$.

For the dephasing model we study, in case of uncontrolled evolution, it is known that non-Markovianity manifests only when super-ohmic spectral densities are considered \cite{discfro}. Note that the emergence of non-Markovian behavior here is directly connected with the decay rate $\gamma_0(t)$ being temporarily negative during the dynamics, which occurs for $s>2$. At this point, we should also mention that although the approach we follow for characterizing non-Markovianity can be interpreted from several different angles based on the information exchange between the system and its environment, our results in fact do not depend on the specific choice of measure of non-Markovianity, since several well-known methods have been shown to qualitatively agree for the considered dephasing model \cite{rivas10,breuer09,luo12,bylicka14}.

\begin{figure}[t]
\centering
\includegraphics[width=0.5\textwidth]{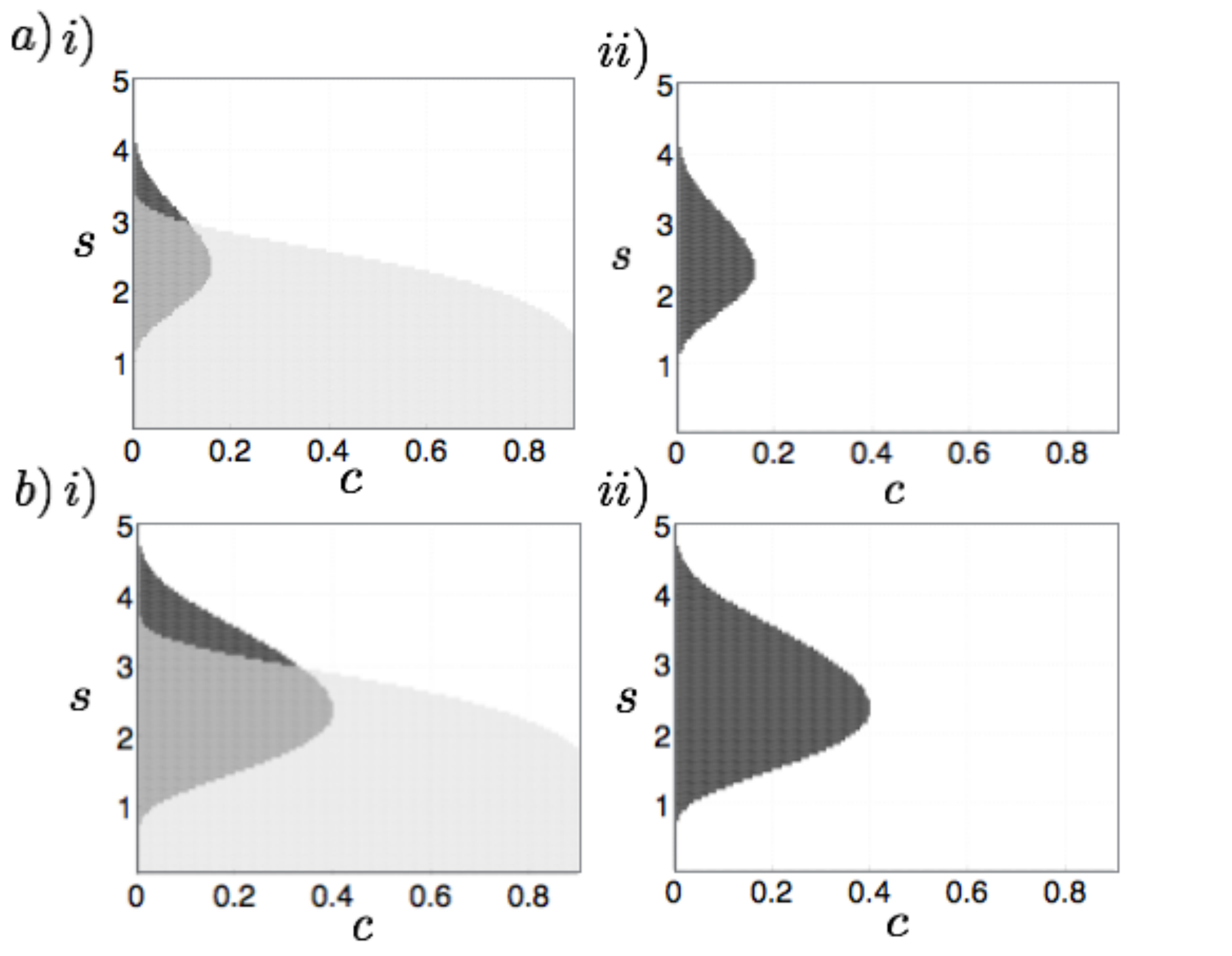}
\caption{(Color online) The black region shows the range of $s$ and $c$ parameters for which the discord is frozen forever for a free system evolving without pulsing. The gray region shows the range of $s$ and $c$ for dynamical decoupled systems with small interval spacing $\omega_c\Delta t=0.3$ (i) and long interval spacing $\omega_c\Delta t=3$ (ii).
The plots in (a) are for the case where both qubits are affected by noise, as the plots in (b) are for single qubit noise. Outside these regions, one will always observe a transition from classical to quantum decoherence and thus no time-invariant discord. The final pulse is applied at $t_f=N_\text{max}\Delta t\leq 25\omega_c^{-1}$ where $N_\text{max}$ is the maximum number of pulses that can be applied within the time interval $0\leq t\leq 25\omega_c^{-1}$. Quantities plotted are dimensionless. }
\label{A1}
\end{figure}

For simplicity, we consider the case of periodic DD pulses, applied at times $t_n=n\Delta t$ with $n=1,2,3,\ldots$. We define two pulse intervals with respect to the onset of non-Markovian dynamics. In more detail, short pulse intervals are defined as $\Delta t<\tilde t$ where $\tilde t=\tan(\pi/s)$ and $\gamma_0(\tilde t)=0$ and conversely for long pulse intervals. Here, $\tilde t$ denotes the first time instant after which the information flow is reversed for the uncontrolled dynamics. It was shown in Ref. \cite{addis} that optimal DD efficiency is always achieved for Markovian regimes. In particular, for long pulse intervals, DD pulsing can lead to rapid decoherence relative to the free dynamics. Indeed, if a pulse occurs during a period of re-coherence, initiated by the free non-Markovian dynamics, information flow will be inverted leading to decoherence. In the light of these results, it becomes natural to relate the findings of Ref. \cite{discinv} on the competition between non-Markovianity and DD to the creation of time-invariant discord. Specifically, we intend to establish a set of necessary conditions for creating time-invariant discord.

\begin{figure*}[t]
\centering
\includegraphics[width=0.85\textwidth]{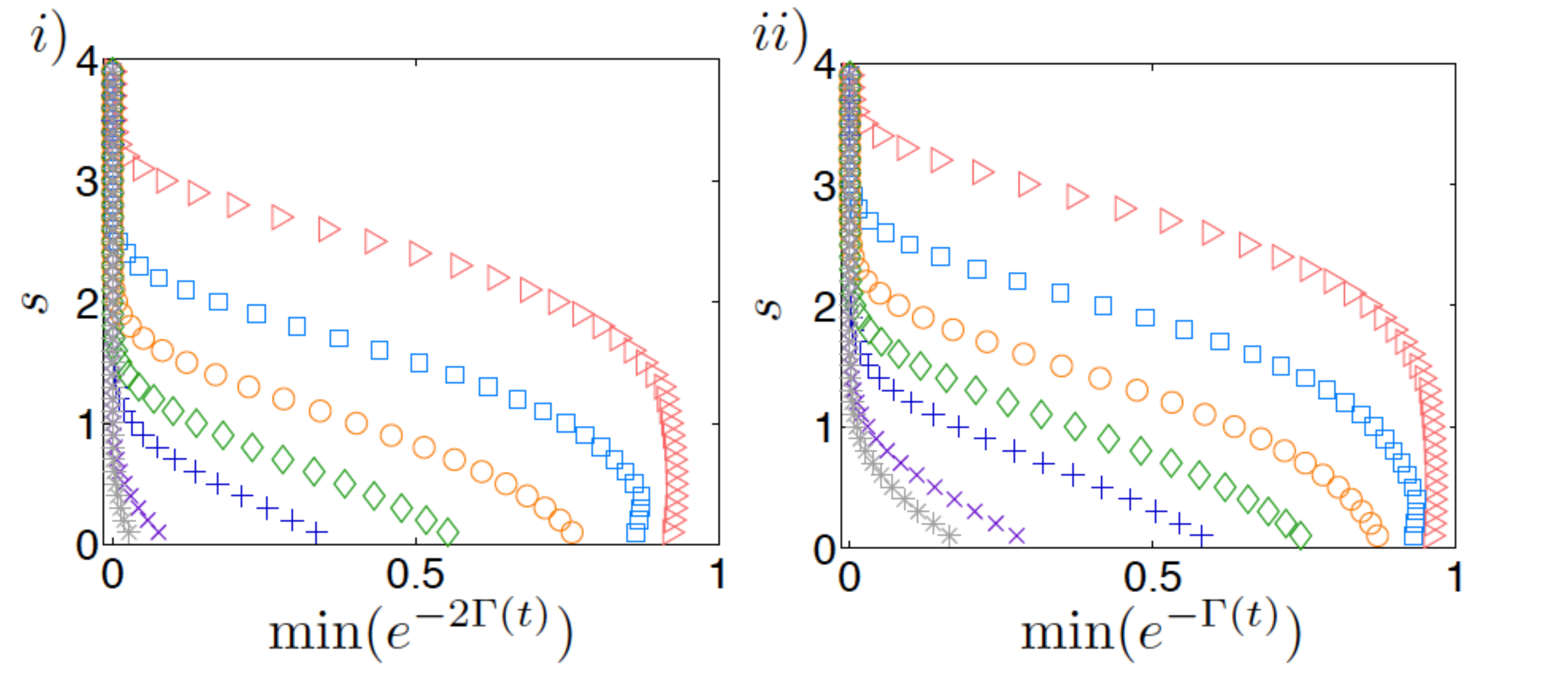}
\caption{(Color online) The boundary between the sudden transition and the time-invariant discord as a function of the Ohmicity parameter $s$ and the minimum value of decoherence factor throughout the time evolution, for two-sided (i) and one-sided (ii) dephasing noise. We show pulse intervals $\omega_c\Delta t=0.3$ (red triangles), $\omega_c\Delta t=0.4$ (blue squares), $\omega_c\Delta t=0.5$ (orange circles), $\omega_c\Delta t=0.6$ (green diamonds), $\omega_c\Delta t=1$ (blue plus signs), $\omega_c\Delta t=1.5$ (purple crosses) and $\omega_c\Delta t=2$ (grey stars). The final pulse is applied at $t_f=N_\text{max}\Delta t\leq 25\omega_c^{-1}$ where $N_\text{max}$ is the maximum number of pulses that can be applied within the time interval $0\leq t\leq 25\omega_c^{-1}$. Quantities plotted are dimensionless. }
\label{A2}
\end{figure*}

As continuously pulsed dynamics in the asymptotic time limit can not be defined analytically or numerically, we focus rather on discord which remains ``time-invariant" within a given experimental time rather than forever. Note that the values of invariant discord are given by $Q=(1+c)\text{log}_2(1+c)/2+(1-c)\text{log}_2(1-c)/2$, thus discord acquires a significant value for larger values of $c$. Fig. 1a (i) shows that, for two-sided local dephasing noise and short pulse intervals (e.g. $\omega_c\Delta t=0.3$), time-invariant discord is created for a wider range of both $c$ and $s$  compared to the unperturbed case.  Interestingly, we see that DD pulses allow for the existence of time-invariant discord also in the sub-Ohmic case ($s<1$), for up to very high values of $c$. Generally, one can see from the plot that increasingly significant values of time-invariant discord (corresponding to large values of $c$) occur for $s<2$. Numerical investigation shows that these conclusions are independent of the specific pulse interval chosen, provided that $\Delta t<\tilde t$.

In the asymptotic long time limit ($t_f\rightarrow\infty$), based on numerical evidence, we conjecture that DD will only create invariant discord for $s\lesssim 1$. In more detail, as no degradation of the coherence is shown to occur within computable times, we assume that in the asymptotic long time limit ($t_f\rightarrow\infty$), coherence is maintained for values close to unity. On the other hand, for $s\geq 1$, as $t_f$ increases, the region of time-invariant discord will decrease as the coherence decays to smaller and smaller values. 

In Fig. 1a (ii), we show that time-invariant discord is always destroyed by DD techniques for large pulse intervals (e.g, $\omega_c\Delta t=3$) in case of two-sided noise. Once again, we have checked numerically that this behaviour characterises the large pulse interval region. We conclude that for short pulse interval DD schemes and Markovian environments, specifically $s<1$, one can create invariant discord for a larger range of initial states compared to relying on non-Markovianity alone as a resource.  For the sake of comparison we show in Fig. 1b (i) and Fig. 1b (ii) exemplary plots illustrating the results of the same analysis when only one of the qubits is affected by dephasing noise. Although we observe similar results to the previous case in general, we note that the regions of invariant discord, in both the uncontrolled and the controlled dynamics cases, become larger. However, we note that whereas the difference in the regions of $s$ and $c$ between the one-sided and two-sided noise is significant in the case of unpulsed non-Markovian dynamics, the difference is less pronounced when we implement the DD pulses.

Let us now investigate the region of parameters $c$ and $s$ giving rise to time-invariant discord for \emph{intermediate} values of pulse interval $\omega_c\Delta t$. The behaviour of the boundary between the sudden transition and the time-invariant discord for different values of the pulse interval is displayed in Fig. 2 (i) and Fig. 2 (ii) for the two-sided and the one-sided local dephasing noise, respectively. Analogously to Fig. 1, we plot the transition boundary, below which, discord is time-invariant and classical correlations decay. Note that decoherence factors are set to unity initially at $t=0$. Therefore, for a fixed $s$, if the initial state parameter $0 \le c\le 1$ is chosen so that it is smaller than the minimum value of the decoherence factor $e^{-\Gamma(t)}$ at any point during the pulsed dynamics, there occurs no sudden transition and consequently discord remains invariant in time at all times. However, when $c>\text{min}(e^{-\Gamma(t)})$ for a fixed Ohmicity parameter $s$, then this means that a sudden transition takes place and discord begins to decay after the critical time instant, which can obtained from the solution of the equation $\text{min}(e^{-\Gamma(t)})=c$. Therefore, Fig. 2 actually demonstrates how the region of parameters $s$ and $c$ creating time-invariant discord are affected as the pulse interval changes between the short and long interval limits. Comparing Fig. 1 and Fig. 2, we can see that when the pulse interval $\omega_c\Delta t$ gets larger, the overlap between the regions of uncontrolled time-invariant discord and the controlled one becomes smaller. Finally, one can also observe that relying only on non-Markovianity as a resource of time-invariant discord becomes more preferable in general, as the pulse interval $\omega_c\Delta t$ increases.

\begin{figure}[t]
\centering
\includegraphics[width=0.51\textwidth]{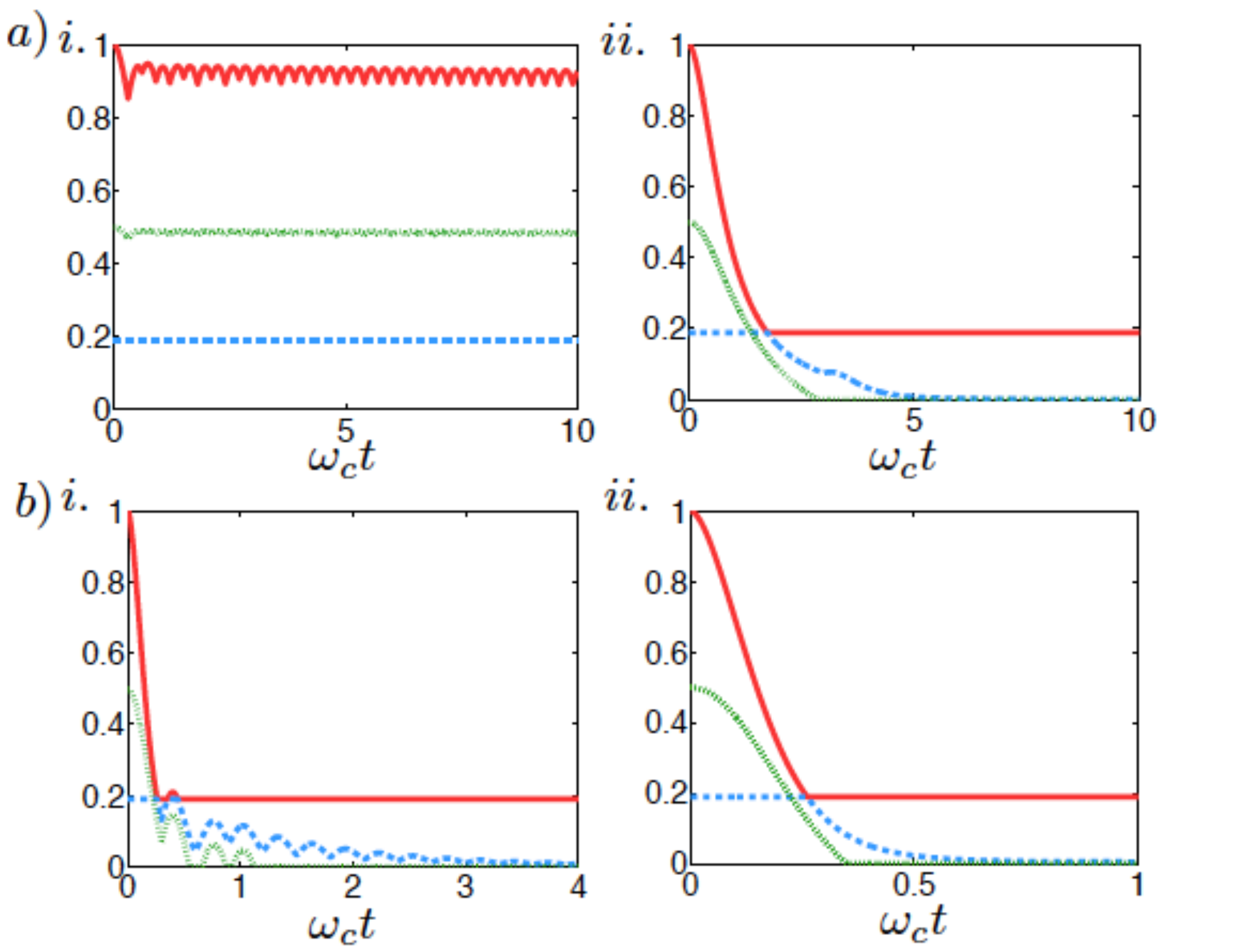}
\caption{(Color online) Dynamics of entanglement (dotted green), classical correlations (solid red) and quantum discord (dashed blue) for i) short pulse spacing $\Delta t=0.3$ and ii) long pulse spacing $\Delta t=3$ (in units of $\omega_c $) and a) Markovian environments (s=1.01) and b) non-Markovian environments (s=4). The plots are for local dephasing noise on one qubit and $c=0.5$. Quantities plotted are dimensionless. }
\label{A3}
\end{figure}

We conclude our study by illustrating, for a specific initial state, i.e. $c=0.5$, how DD pulses affect the sudden transition and time-invariant discord dynamics for small and large pulse intervals with Markovian/non-Markovian regimes. Moreover, for completeness, we compare the behaviour of discord and classical correlations to the dynamics of entanglement as measured by concurrence:
\eq
C=\frac{1}{2}\text{max}\{0,|e^{-\Gamma(t)}(1-c)|-1-c,|e^{-\Gamma(t)}(1+c)|-1+c\},
\eeq
where we assume that one of the qubits is locally interacting with the environment. For short pulse intervals ($\Delta t<\tan(\pi/s)$), and Markovian dynamics ($s<2$), both classical correlations and entanglement are well preserved while discord remains invariant at a non-zero value for the experimental time interval considered, as shown clearly in Fig. 3a (i). On the other hand, even if one of these conditions is not satisfied, time-invariant discord can vanish due to the sudden transition between classical and quantum decoherence, and entanglement decays to zero. Quite intuitively, in order to preserve entanglement through dynamical decoupling, a pulse needs to occur before entanglement decays to zero (see Fig. 3 (ii)).

We note that, as one can see from Fig. 2, for $c=0.5$ time-invariant discord occurs only for smaller pulse spacings ($\omega_c\Delta t\lessapprox 0.6$). Moreover, only for very small pulse spacing ($\omega_c\Delta t\approx 0.3$), can invariant discord be achieved for non-Markovian values (see Fig. 2 (i), red curve). Once discord become invariant, the value depends only on the initial state through the parameter $c$.

Generally, in the Zeno regime and for divisible (Markovian) dynamics, dynamical decoupling protocols preserve quantum properties with increased efficiency for increasingly small pulse spacing \cite{Zeno2}-\cite{Zeno1}. The same holds for indivisible dynamics (non-Markovian) provided that $\tilde t>\Delta t$. In the short pulsed regime, it has been shown that as $s$ increases, coherence is preserved less efficiently through DD protocols, revealing the detrimental influence of non-Markovian dynamics \cite{addis}. For $\tilde t<\Delta t$, non-Markovian dynamics contribute to revivals in the dynamics and hence no general conclusion can be drawn.  

\section{Conclusions}

The persistence of quantum correlations in presence of noise induced by the environment is a remarkable phenomenon which may be crucial for those quantum information tasks using such correlations as a resource. 
In this paper we have studied the intriguing phenomenon of time-invariant discord with the aim of understanding whether the presence of DD pulses might lead to a wider range of parameters for which it occurs and if, in this way, one can obtain higher values of ``protected" discord.

We have performed an extensive analytical and numerical study exploiting an approach developed in Ref.  \cite{addis} in a one-qubit scenario. While in the unpulsed regime time-invariant discord mainly occurs in the non-Markovian regime, our work proves that DD pulses allow for this phenomenon to take place also in the Markovian regime. In fact, our results show that, in the pulsed regime, the optimal conditions to achieve significant values of time-invariant discord persisting for the whole experimental time scales are obtained for  Markovian reservoirs ($s<2$) in the the short-pulse regime. Based on numerical analysis, we conjecture that for discord to remain invariant to non-zero values forever, i.e. in the limit $t_f \rightarrow \infty$, the Ohmcity parameter is further restricted to $s<1$. On the contrary, large pulse spacing and free non-Markovian dynamics are detriminental to the creation of time-invariant discord, in the pulsed scenario. 

Reservoir engineering is one of the most promising tools for fighting the loss of quantum properties, with prominent applications to realistic quantum technologies. Here we have considered two types of reservoir engineering, one based on a modification of the reservoir spectral density (and hence the Ohmicity parameter), leading to a change in the Markovian character of the dynamics, and the second one exploiting pulses applied on the system (which, in turn, can also be seen as some sort of filtering of the spectral density). 
Our results show the optimal strategy to use, depending on the system-environment characteristic parameters.

\section{Acknowledgements}
C.A. acknowledges financial support from the EPSRC (UK) via the Doctoral Training Centre in Condensed Matter Physics under grant number EP/G03673X/1. G. K. is grateful to Sao Paulo Research Foundation (FAPESP) for the BEPE postdoctoral fellowship given under grant number 2014/20941-7. S. M. acknowledges financial support from the EU Collaborative project QuProCS (Grant Agreement 641277), the Magnus Ehrnrooth Foundation, and the Academy of Finland (Project no. 287750).

\end{document}